\documentstyle[12pt]{article}
\input epsf
\begin{document}

\newcommand{\pho}{\tilde{\gamma}}
\newcommand{\gl}{\tilde{g}}
\newcommand{\sneu}{\tilde{\nu}}
\newcommand{\sigmin}{\sigma_{min}}
\newcommand{\st}{\tilde{t}}
\newcommand{\sq}{\tilde{q}}
\newcommand{\se}{\tilde{e}}
\newcommand{\ch}{\chi^{\pm}}
\newcommand{\neut}{\chi^{0}}
\newcommand{\gsi}{\,\raisebox{-0.13cm}{$\stackrel{\textstyle>}
{\textstyle\sim}$}\,}
\newcommand{\lsi}{\,\raisebox{-0.13cm}{$\stackrel{\textstyle<}
{\textstyle\sim}$}\,}
\newcommand{\be}{\begin{equation}} \newcommand{\ee}{\end {equation}}

\rightline{RU-97-20} 
\rightline{hep-ph/9706393}
\rightline{June, 1997} 
\baselineskip=18pt \vskip 0.7in 
\begin{center} {\bf \LARGE $e^+ e^-$ Cross Section and Exclusion  \\
of Massless Electroweak Gauginos}\\ 
\vspace*{0.9in} 
{\large Glennys R. Farrar}\footnote{Research supported in part by
NSF-PHY-94-23002} \\ 
\vspace{.1in} 
{\it Department of Physics and Astronomy \\ Rutgers
University, Piscataway, NJ 08855, USA}\\ 
\end{center} 
\vspace*{0.2in}
\vskip 0.3in 

{\bf Abstract:} Measurements of the total hadronic cross section in
$e^+ e^-$ annihilation are shown to be capable of severely limiting the
possibility that gauginos have negligible tree level masses.  A combined
analysis of 1997 and earlier LEP data, considering simultaneously conventional
SUSY signatures and purely hadronic final states, should achieve a 95\% cl
sensitivity to the case that the SU(2) and U(1) gauginos are massless.
If integrated luminosity targets are achieved, it should also be possible
to exclude the case that the wino or wino and gluino are light while the
bino is heavy, except possibly for a small region of $\mu,~ tan \beta$.
The analysis applies whether or not R-parity is conserved, and can
also be used to reduce the model-dependence of conventional SUSY searches.
 
\thispagestyle{empty} 
\newpage 
\addtocounter{page}{-1}
\newpage

An important prediction of SUSY breaking scenarios which give small or
negligible mass, $m_2$, to the SU(2) gauginos (winos) is that the lighter
physical chargino has mass $\lsi m_W$.\footnote{ $ 2 M_{\ch}^2 = \mu^2 +
2 m_W^2 \pm \sqrt{\mu^4 + 4 m_W^4 cos^2 2 \beta + 4 m_W^2 \mu^2}$ at tree
level.}
If the U(1) gaugino (bino) also has negligible mass, $m_1$, then the
lightest neutralino is a nearly massless photino.  When gaugino masses
are negligible at tree level, radiative corrections give a gluino mass
of $\sim 100$ MeV and a somewhat larger photino mass, depending on
parameters\cite{radmass}.  The photino can account for
the observed dark matter\cite{cosmo} if it is lighter than the
gravitino and $R$-parity is conserved.  

\begin{figure}
\epsfxsize=\hsize
\epsffile{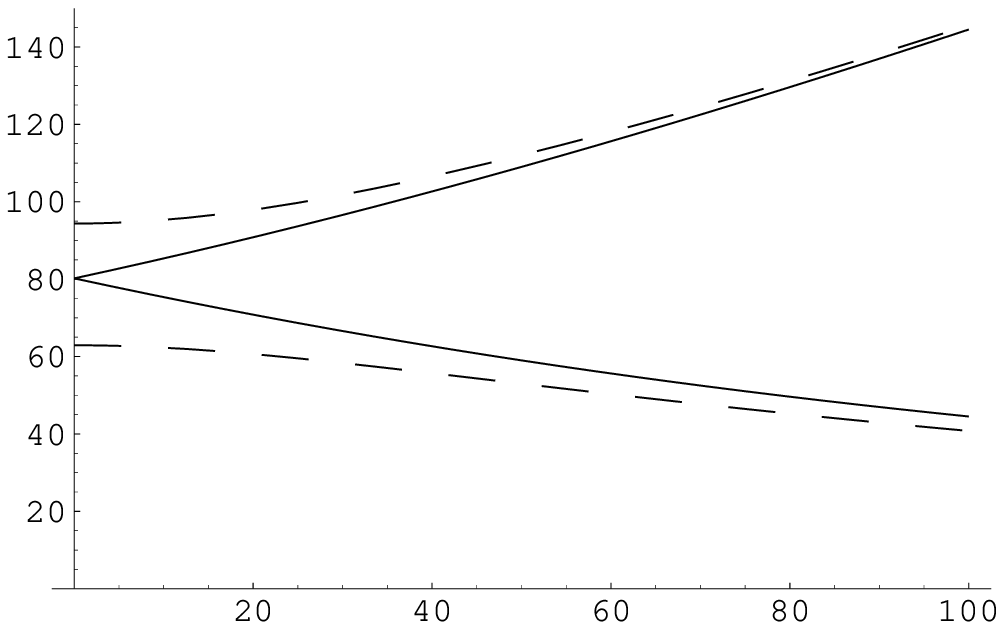}
\caption{Chargino masses versus $\mu$ in GeV, for $tan \beta = 1$
(solid) and 1.5 (dashed).}
\label{fig:chmasses}
\end{figure}

\begin{figure}
\epsfxsize=\hsize
\epsffile{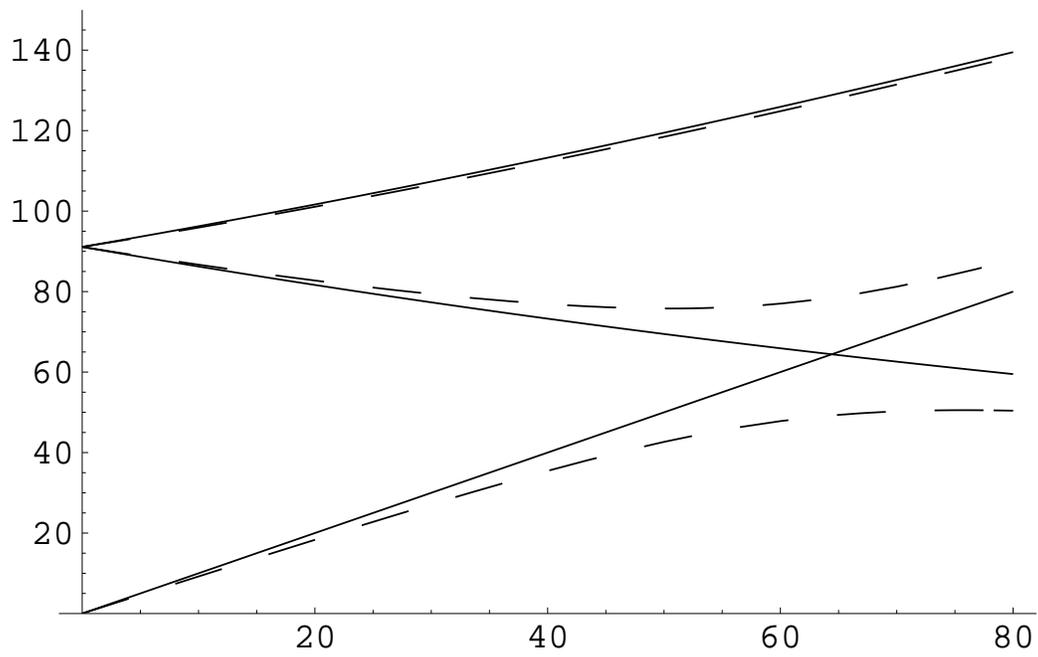}
\caption{Masses of $\chi^0_2,~\chi^0_2$ and $\chi^0_4$ versus $\mu$
(GeV), for $tan \beta = 1$ (solid) and 1.5 (dashed).}
\label{fig:nmasses}
\end{figure}

In such scenarios, the cross section for production of charginos and
neutralinos is determined essentially completely by $\mu$, the
higgsino mixing parameter, $ ~tan \beta \equiv v_U/v_D$, and
$m(\tilde{\nu_e})$.  $\mu$ and $tan \beta$ determine the masses and
couplings of the charginos and neutralinos;  the sensitivity to
$m(\tilde{\nu_e})$ is due to destructive interference between the
$s$-channel poles and $t$-channel $\tilde{\nu_e}$ exchange.  Figures
\ref{fig:chmasses} and \ref{fig:nmasses} show the chargino and neutralino
masses as a function of $\mu$, for several values of $tan \beta$ and
$m_1 = m_2 = 0$.  It is easy to see analytically that when $m_1$ and
$m_2$ are negligible and $\mu$ obeys $ \mu << m_Z$, the second
lightest neutralino is mainly higgsino and has mass approximately
$\mu$.   

The important feature to extract from Figs. \ref{fig:chmasses} and
\ref{fig:nmasses} is that $m(\chi^0_2)+m(\chi^0_3)$ is smallest at
low $\mu$ while $2 m(\chi^\pm_1)$ is smallest at large $\mu$.
Therefore for any value of $\mu $ either neutralino or chargino
production is kinematically favorable.  The purpose of this paper is
to point out that the entire parameter space of any scenario with
approximately massless electroweak gauginos leads to rates for either
$e^+ e^- \rightarrow \chi^0_2 \chi^0_{3,4}$ or $e^+ e^- \rightarrow
\chi_i^+ \chi_j^-$ which are accessible with existing LEP data.  Scenarios
with a heavy bino but approximately massless wino are potentially also 
fully accessible with the planned 183 GeV and 192 GeV data acquisition.  

Depending on the relative mass of squark and $W$, whether the
gluino is light or not, whether $R$-parity is violated and if so how,
chargino and neutralino ({\it ino}) pair production will contribute mainly
to the hadronic cross section or to conventional SUSY signatures (missing
energy, plus leptons and/or multiple jets\footnote{Or, in a window around
$m_1 \approx m_2 \approx \mu \approx 0$, to apparent excess $W^\pm$ pair
production\cite{fpt}.}). However in all cases, some
anomaly in comparison to SM predictions will be experimentally observable.  

If the gluino is heavy and short-lived and $R$-parity is conserved,
the standard phenomenology with missing energy applies\cite{ff} so the
gluino mass must be greater than 154 GeV\cite{pdg96}.  In this scenario
charginos and neutralinos are too light to decay via a gluino. 
For their conventional chargino and neutralino analysis, LEP experiments
place a limit on excess events of the types  
\begin{itemize}
\item  $E_T^{\rm miss}$ and hadronic activity in four or more jets.
\item  $E_T^{\rm miss}$, two or more jets, and a charged lepton.
\item  $E_T^{\rm miss}$ and two oppositely charged leptons, possibly of
different flavor.
\end{itemize} 
These signatures are sensitive to the chargino decays 
\be
\chi^\pm \rightarrow [\chi^0 W^\pm ]\rightarrow \chi^0 f \bar{f};~~
\chi^\pm \rightarrow [\tilde{\nu_l} l~ {\rm or}~ \tilde{l} \nu_l]
\rightarrow \chi^0 l \nu_l
\label{convmodes}
\ee
and the analogous neutralino decays.

When these are the only decay modes, $m_1,~m_2 \approx 0$ is already
excluded\cite{opal:chneut}.  For small bino and wino mass, the mass gap between
the neutralinos $\chi_2^0$ and $\chi_1^0$ is about $\mu$.  Thus when $\mu$
is larger than a few GeV the experimental acceptance for the conventional
decay modes (\ref{convmodes}) is good\cite{aleph:susy133,opal:chneut} and
the sensitivity is sufficient to exclude $m_1,~m_2 \approx 0$.  When
$\mu,~m_1,~m_2$ are all of order GeV's, the analysis of Feng, Polonsky, and
Thomas\cite{fpt} applies. This window has now been excluded, again assuming
the modes (\ref{convmodes}) are the only important decay channels\cite{opal:chneut}.

However other decay modes are possible, and indeed may dominate, if the
gluino is light\footnote{Light, long-lived gluinos escape detection by
conventional means such as beam dump experiments or limits on missing
energy\cite{ltgl}.  Most of the interesting range of mass and lifetime
has yet to be directly explored experimentally\cite{f:95,f:104}.
Nor are indirect tests capbable of excluding light gluinos due to QCD
uncertainties.  For instance, when the theoretical uncertainty
associated with scale dependence and hadronization models are included
as in other experiments, the recent ALEPH limit $m_{\tilde{g}} > 6.3$ 
GeV\cite{aleph:lg} becomes instead compatible with a massless
gluino\cite{f:118}.  Also, the systematic uncertainties in $\alpha_s$
at both low and high $Q^2$ are still too large to use the $Q^2$ 
dependence to probe the light gluino possibility.  Ref.
\cite{shifman96} shows that the uncertainty in $\alpha_s(m_\tau)$ is
much larger than previously assumed so that the $Q^2$ dependence of
$\alpha_s$ only slightly (70 \% cl) disfavors a light
gluino\cite{csikor_fodor}. {\it Note Added:}  See \cite{f:119} for a
review of these and other recent experimental constraints.} or $R$-parity
is violated.  In the light gluino case these are: 
\be
\chi^\pm \rightarrow [\tilde{q} \bar{q}~ {\rm or}~ \tilde{\bar{q}} q ]
\rightarrow \gl \bar{q} q.
\label{ltglmodes}
\ee
In the $R$-parity violating case the intermediate squark could
produce quark pairs\cite{rparviol}.  High energy quarks and gluinos
hadronize to produce jets\cite{ltgl}, so squarks decaying to quark and
gluino or to quark pairs produce two jets, with little missing
energy\cite{f:105,rparviol}.  $R$-parity and lepton number violating decays
$\sq \rightarrow q + l$ are included in the hadronic event sample,
as long as the missing energy is small.

The branching fractions to the various final states (\ref{convmodes}),
(\ref{ltglmodes}) depend critically on the relative masses of {\it ino}, sneutrino
or slepton, squarks and $W$. Two-body on-shell decays dominate three body
decays, except at the extreme edge of phase space.  At present, selectron
and sneutrino mass limits (about 60 GeV and 45 GeV respectively\cite{pdg96})
are compatible with the possibility that the two-body decays $\chi^{\pm 0}
\rightarrow \tilde{\nu} l,~\tilde{l} \nu,~\tilde{l} l,~\tilde{\nu} \nu$ are
kinematically allowed.  Limits on squark masses\footnote{Up and down squark
masses must either be larger than $\sim 600$ GeV or in the range $\sim 50
- 130$ GeV. The lower limit is to ensure compatibility with the $Z$ 
width\cite{sqlimZ} and the upper limit to avoid dijet mass peaks above QCD
background at the FNAL collider\cite{sqlimppbar,f:105}.} are compatible with
the possibility that on-shell $\chi^{\pm 0} \rightarrow \tilde{q} q$ is
kinematically allowed.  If both types of two-body decays are allowed,
the color multiciplicity of the $q \tilde{q}$ final state causes
hadronic decays to dominate, as for $W$ decay.  If no two-body decay
modes are allowed\cite{ino},  
\be
\Gamma(\chi^\pm \rightarrow q \bar{q} \gl)/\Gamma(\chi^\pm
\rightarrow \chi^0_1 W^\pm) \sim 5 (\frac{m_W}{m(\tilde{q})})^4.
\ee
Thus, for a substantial part of the allowed squark mass range, the
hadronic channels (\ref{ltglmodes}) dominate chargino and neutralino
decay. 

While the decay modes (\ref{ltglmodes}) preferentially populate the multijet
($n \geq 4$) sample, the best way to put limits on {\it ino}'s which
decay through these modes is to use limits on an excess in the {\it
total} hadronic cross section.  The total hadronic cross section is far less
sensitive to QCD uncertainties than are individual $n$-jet cross 
sections.  QCD predictions for experimental observables which involve
definition of jets require resummation of the large logarithms of the
small parameter $y_{\rm cut}$ used to define the jets.  Furthermore
QCD predictions for $n-$jet cross sections are very sensitive to
$\alpha_s$ and its running, to the hadronization model, and to the
scale dependence from truncation of the perturbation series.  Thus
being able to use the total cross section greatly reduces the theoretical
uncertainty of the SM prediction.

When one of the two {\it ino}s decays via (\ref{ltglmodes}) and the
other via (\ref{convmodes}), the events would often be picked up by the 
conventional search, with somewhat modified efficiency.  However let us
begin by analyzing the simplest limit, when {\it ino}s other than the
lightest neutralino decay hadronically 100\% of the time.  This
phenomenology has not previously been explored.  The treatment of the more
general case of decay both through (\ref{convmodes}) and (\ref{ltglmodes})
is discussed afterward. 

Our experimental constraints come from the very high statistics
measurement of the total hadronic cross section on the $Z^0$ peak, and
lower statistics measurements at higher energy.  The PDG value
for the total hadronic cross section on the $Z^0$ peak is
$\sigma^0_{\rm had} = 41.54 \pm 0.14$ nb\cite{pdg96}. 
Assuming no deviation from the SM prediction, this error corresponds
to a 95\% cl upper limit on any new physics contribution of $1.64 \times
140$ pb = 230 pb.  Due to higher order QCD corrections to the $e^+ e^-
\rightarrow q \bar{q}$ rate, the theoretical prediction for
$\sigma^0_{\rm had}$ has an irreducible fractional uncertainty of
$\frac{\delta(\alpha_s)}{\pi} \approx 0.003$, but this produces an
uncertainty of less than $\pm 0.1$ pb in $\sigma^0_{\rm
had}$\footnote{The best determination of $\alpha_s(m_Z)$ which does
not employ $\sigma^0_{\rm had}$ makes use of event shape variables,
but results of this method display a large dispersion\cite{burrows:warsaw}.}.  

According to OPAL results\cite{opal:pn280} uncertainties in the cross
sections of some 2 pb per experiment can be obtained with the current
data collected in the high energy running of LEP.  At present, the
error in estimating the feedthrough for the $s'/s $ cut dominates the
systematic error.  The systematic uncertainty in the $W^+ W^-$
contribution to the hadronic cross section, which must be included to
obtain the SM   prediction, is negligible at this level.  One can
expect that when all four LEP experiments combine their results, the
final statistical and systematic error can be reduced by a factor of
two from the preliminary OPAL values.  Then if no excess signal is
observed, the 95\% cl upper limits will be about 2 pb or slightly
better\footnote{I am grateful to P. Mattig for help with this estimate
and those below.}.  With 100 pb$^{-1}$ per experiment at 183 GeV, 
one can anticipate an error bar of $\approx 0.5$ pb, and thus a
$\approx 0.8$ pb 95\% cl upper limit.   

For a given $\mu,~tan \beta$, find the sneutrino mass (consistent with
experimental limits) which minimizes the total cross section for $e^+ e^-
\rightarrow (\chi_i^0 \chi_j^0 ~+~ \chi_i^+ \chi_j^- )$\footnote{Throughout,
cross sections include initial state radiation and the cut $s'/s \geq 0.8$ 
as in the OPAL analysis.  They are calculated using the programs of
M. Mangano, G. Ridolfi et al\cite{mlm} with minor modifications.},
summed over all {\it ino} species $i,j$ except the lightest neutralino.  Denote
this cross section $\sigmin(\mu,~tan \beta)$, or $\sigmin$ when minimizing
over $\mu,~tan \beta$ as well.  At 172 GeV, $\sigmin = 2.0$ pb; it is realized
for $\mu = 0,~tan \beta = 1.2$, and $m_{\tilde{\nu_e}} \approx 85$ GeV.
If a combined analysis of the existing LEP 172 GeV data can produce a 95
\% cl upper limit on a deviation from the SM hadronic cross section of 2
pb or lower, the all-gauginos-massless scenario would be excluded, assuming
purely hadronic decay of the {\it ino}s. If some hint of a signal is seen
in the existing data, it should be unambiguously confirmed or excluded with
a high statistics run at 183 GeV, where the cross section for $e^+ e^- \rightarrow
(\chi^0 \chi^0 ~+~ \chi^+ \chi^- )$ with $\mu = 0,~ tan \beta = 1.2$ is 4
pb. Indeed, at 183 GeV $\sigmin = 2$ pb, for $\mu=0,~tan \beta = 1.4$; this
is significantly above the 95 \% cl sensitivity estimated above.   

The generic case with decay to both hadronic and conventional-SUSY final
states can be approached in two ways.  The ``brute force" method is to
generalize the usual SUSY analysis to allow for the possibility of
squarks decaying hadronically\footnote{And decaying to lepton and quark, if
it is desired to allow for lepton-number-$R$-parity violation.} and then
compare the predicted to observed number of events in suitable classes of
events, including purely hadronic final states as well as those used in
conventional SUSY analyses. 
This analysis is quite complex, because the relative number of events in
the various samples will depend on squark masses.  Thus the limits will depend
on having assumed some squark mass spectrum.  Including the possibility of
two body decays to sneutrinos, $R$-parity violation, etc, makes the analysis
even more complicated and model dependent.

A more elegant approach, if feasible, would be to put limits on the excess
${\it total}$ $e^+ e^-$ cross section, including purely hadronic as well
as conventional SUSY channels and mixed channels.  This would reduce
the model dependence of the analysis even in the case of purely conventional
decays.  For instance it removes the model dependent systematic uncertainty
associated with the efficiency for a signal event to pass the cuts (e.g.,
missing energy) defining a particular class of events.  In
particular, the limit of small splitting between {\it ino} mass eigenstates
is no more difficult to treat than large mass splitting.  However this approach
requires computing the cross section for ``radiative returns" ($e^+ e^-
\rightarrow \gamma 's + Z^0$) with a systematic uncertainty which is small
compared to the desired $\approx 2$ pb limit on excess cross section.  If
this is possible, the approach suggested here will provide a new and
potentially more powerful strategy for {\it ino} searches, which is applicable
whether or not gauginos are light and only depends on an increase in
total cross section which is larger than the systematic (and statistical)
uncertainties. 

Finally, an intermediate approach can work if limits on the excess cross
section in conventional SUSY final states ($\sigma_{min}^{conv}$) and in
purely hadronic final states ($\sigma_{min}^{had}$) are good enough. If
the linear combination $b \sigma_{min}^{had} + (1-b) \sigma_{min}^{conv}$
is lower than $ \sigmin(\mu,~tan \beta) $ for any $b$ satisfying $0 \leq
b \leq 1$, that value of $\mu,~tan \beta$ is ruled
out if $m_1,~m_2 \approx 0$.  At present, the limits are not good enough
to exclude all values of $\mu,~tan \beta$ except for $b \approx 0$ or 1.
However the anticipated integrated luminosity of the next two LEP runs,
should put this test within reach.  Assuming no anomaly is found, this would
exclude $m_1,~m_2 \approx 0$ for any value of squark, sneutrino, and gluino
mass and without any assumption about the size of $R$-parity violating
couplings. 

It is interesting to consider the possibility that some but not
all of the gaugino masses are negligible.  This does not occur in
the simplest gravity-mediated SUSY breaking, but can arise in more
general SUGRA models.  Moreover when SUSY breaking is communicated to
the visible sector through particles with standard model gauge
interactions, some, none or all of the gauginos obtain masses at
leading order, depending on the gauge quantum numbers of the
messengers.  For instance binos, but not gluinos and winos, have large
masses if the messengers have only hypercharge interactions\cite{mn}.
One can also consider the possibility of small $m_1$ and $m_3$ but large 
$m_2$, or $m_3 << m_1$ and $m_2$\cite{raby:gmsb}.  

\begin{figure}
\epsfxsize=\hsize
\epsffile{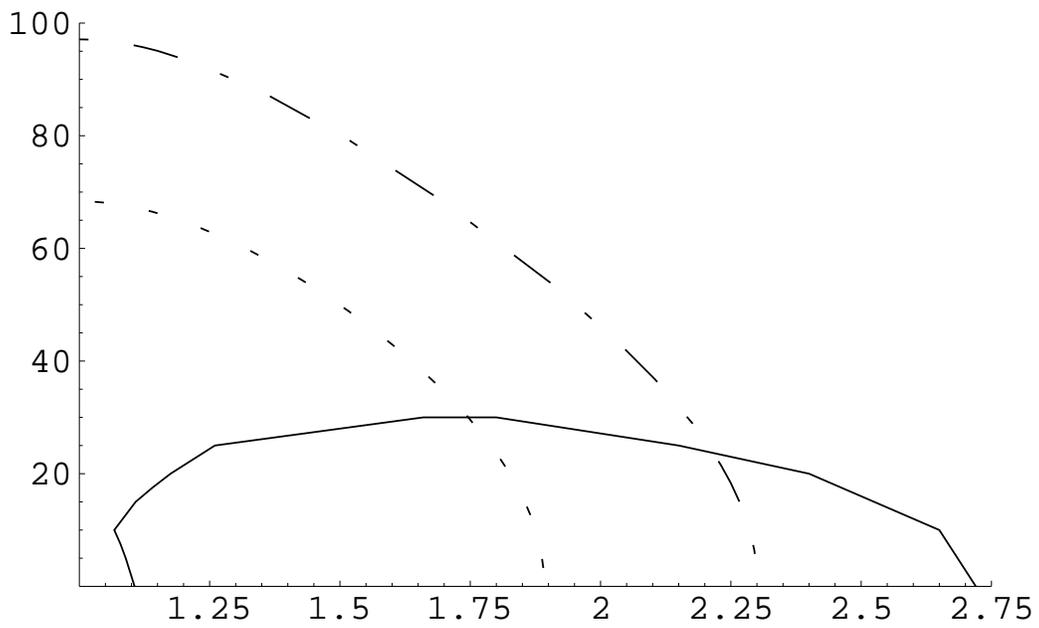}
\caption{Curves of constant $m(\chi^\pm)$: 45 GeV (dash-dot), 53
GeV (dot).  Chargino production alone is inconsistent with the $Z^0$
width in the region to the right of the dash-dot curve and with the
anticipated 172 GeV sensitivity limit above the solid curve.}
\label{fig:chlims}
\end{figure}

The cross section limits anticipated here would severely restrict
scenarios in which large $m_1$ pushes neutralinos out of reach, yet
charginos are pair produced because $m_2 \approx 0$.  This is
illustrated in Fig. \ref{fig:chlims}.  The area below and to the left
of the dash-dot curve is the region in the $\mu, ~tan \beta$ plane
which is compatible with the requirement that $\Gamma(Z^0 \rightarrow
\chi^+ \chi^-) < 230$ pb, i.e., $m(\chi^\pm_1) = 45$ GeV.  The area
below the solid curve in Fig. \ref{fig:chlims} is the $\mu,~ tan
\beta$ region which would be consisent with a 1.9 pb limit on
$\sigma_{had}$ at 172 GeV, for $b=1$.  However one can do better than this.
From Fig. \ref{fig:chmasses} one sees that $\chi_1^+ \chi_1^-$
production is kinematically allowed even at 133 GeV for the larger
$tan \beta$ values in the allowed region of Fig. \ref{fig:chlims}.
Furthermore, the value of $m(\tilde{\nu_e}) $ which minimizes the
cross section is a relatively strong function of $E_{cm}$.  Thus
demanding that predicted cross sections are above the experimental
limits at all three energies, for a common value of
$m(\tilde{\nu_e})$, will improve the limits indicated in
Fig. \ref{fig:chlims}.  

Overall, the previous discussion indicates that analysis of data at
and below 172 GeV will allow the parameter space for $m_2 \approx 0$
to be limited to $\mu \lsi 30~ {\rm GeV},~ 1.1 \lsi tan \beta \lsi 2$,
for $b=1$.  At 183 GeV, the minimal total cross section for $e^+ e^-
\rightarrow \chi^\pm \chi^\mp $ is 0.6 pb, for $\mu \lsi 1$ GeV, $tan \beta
= 1.4$, and $m(\tilde{\nu_e}) \approx 85$ GeV.  Combining with the lower
energy limits, this is on the borderline of being excludable with 183
GeV data, again for $b=1$.

The dotted curve in Fig. \ref{fig:chlims} corresponds to $m(\chi^\pm_1)
= 53$ GeV.  This is well within the allowed region anticipated for existing
data.  Thus the possibility that the anomalous 4-jet events observed
by ALEPH\cite{aleph:4j} are due to chargino pair production\cite{ino}
cannot be ruled out with present data\footnote{{\it Note added:} No anomaly
in 4-jet events was observed in a repeat run at 130 and 133 GeV with
total luminosity of about 5 $pb ^{-1}$ per experiment.  M. Schmitt,
private communication.}.   

If it is established that $m_2$ is not small, cross section
predictions become much less constrained and ruling out the small
$m_1$ but large $m_2$ case may be difficult.  One interesting, albeit 
fine-tuned possibility is the case that both $m_1$ and $m_2$ are
large, but related in such a way that the lightest neutralino mass is
about 1 GeV.  This is of interest because the $\chi^0_1$ may then be a
good dark matter candidate if it is the LSP and $R$-parity is
conserved\cite{cosmo}.  Another interesting phenomenological
possibility is $m_1 \approx m_2 > \mu$, so that $\chi^0_1$ is
higgsino-like while $\chi^0_2$ is photino-like.  In this case
$\chi^0_2$'s produced in selectron decay via $ \chi^0_2 \rightarrow
\chi^0_1 + \gamma$\cite{gordy}.  

In performing the analyses proposed here, the changes in efficiency
for detecting the excess events predicted in the light gaugino
scenario ought to be modeled for completeness.  It should be straightforward
to make the (presumably small) correction coming from the fact that QCD 
produces dominantly 2- and 3- jet events whereas the excess hadronic
events would have $\geq 4$ jets. Gluino jets may also be ``fatter''
than a typical light quark jet, but this should make little difference
to the efficiency. A more subtle complication is that the $R$-hadron
in each gluino jet, typically the gluino-gluon bound state denoted
$R^0$, may decay (dominantly to $\pi^+ \pi^- \pho $\cite{f:104})
before the hadronic calorimeter.  The corresponding $<p^{\rm miss}>$
carried by each photino would be $\lsi p(R^0)/2$.  (We will see below that
$E(R^0)\approx p(R^0)$.)  We can get a rough 
upper limit on $<p(R^0)>$ as follows. Since the $R^0$ 
has light constituents, its production should be softer than is
typical for the charm quark, so we take $<x_{R^0}> ~<
0.4$.\footnote{A crude estimate rather than upper limit would be
$<x_{R^0}> \approx <x_{K}>$.  I am grateful to P. Nason for a
helpful discussion on this point.}  For pair produced {\it ino}s
with mass $\approx E_{cm}/2$, the momentum of the gluino jet is
typically 1/3-1/2 of the {\it ino} mass, giving $<p(R^0)> \lsi 0.4
\times (\frac{1}{6} - \frac{1}{4}) E_{cm} \sim 15 $ GeV for each
gluino jet. Hence if the $R^0$ lifetime is shorter than $\sim 3
~10^{-10}$ sec\footnote{See ref. \cite{f:95} for limits on the $R^0$
lifetime.} it will on average decay before reaching the hadronic
calorimeter and the photino energy (bounded by $\approx 7$ GeV by the
reasoning above) would be lost.  This effect reduces the probability
for the event to pass the $s'/s$ cut and should be modeled more
carefully in a final analysis.  See \cite{ktev:lg} for experimental
constraints on $R^0$ decay.

If some excess hadronic cross section is observed and is due to
hadronically decaying {\it ino}s, the {\it ino}-containing events can be
identified as follows.  They will have a large jet multiplicity:  in
principle 6 jets, however if the intermediate squark is close in mass
to the parent {\it ino}, it may be difficult to resolve all 6 jets.  Two of
the jets (not the soft ones) will be gluinos.  If an $R^0$ decays in
the detector\cite{f:104}, that jet will contain a ``vee'' from the
decay $R^0 \rightarrow \pi^+ \pi^- \pho$.  This can in principle be
distinguished from the decay of a kaon by the missing $p_t$ carried by
the photino and possibly by the invariant mass of the $\pi^+ \pi^-$
pair\cite{f:104}.  Even if the $R^0$ lifetime is such that it rarely
decays in the detector, gluino jets are distinguishable from ordinary
quark jets by having $Q=0$ and possibly looking more gluon-like than
expected for quark jets.  Reconstructing the invariant masses of
systems of jets should in principle produce mass peaks corresponding
to the {\it ino} masses, but it is not obvious that kinematic
reconstruction of jets in high-multiplicity events can be accurate
enough for this to be practical. 

To summarize:  If gluinos are light or $R$-parity is violated, the
phenomenology of charginos and neutralinos can differ dramatically from the
conventional scenario.  {\it Ino} final states can be purely hadronic, with
negligible missing energy.  Nevertheless, we have seen that by combining
limits on {\it both} conventional modes and excess hadronic cross section
the entire parameter space for models with light electroweak gauginos can
be explored, independent of their decay mode. 

{\it Note added:} OPAL has now implemented the analysis we have 
proposed\cite{opal:sighad}.  As anticipated by the estimates given above,
combining all their 130-172 GeV data allows them to exclude the
possibility that both bino and wino are light for any values of $\mu,~tan
\beta, m_{\sneu}$, assuming {\it ino}s decay purely hadronically ($b=1$).  The
integrated luminosity of that data set is not sufficient to extend the analysis
to arbitrary $b$, except over a portion of $\mu,~tan \beta$ space.
The possibilty of a heavy bino but light wino is only viable for $b=1$ in
a small range of $\mu,~tan \beta$, roughly consistent with our estimate shown
in Fig. \ref{fig:chlims}.  The extention of the analysis to arbitrary $b$,
conceivably also for arbitrary $m_2$, may be possible in the near future.

Until the possibility of a light gluino or $R$-parity violation is definitively
excluded, experimental limits on squarks and {\it ino}s should routinely
address the possibility that some fraction of their final states are purely
hadronic.  This will ensure limits with the greatest possible range of validity.

It was also noted that if the systematic uncertainty on the standard
model prediction for the {\it total} $e^+ e^-$ cross section (including
radiative returns) can be reduced to a low enough level, less
model-dependent limits on new physics should be possible.  This is
because assumptions about final states and efficiency of certain cuts
are irrelevant to determining the total production rate of the new
particles.

{\bf Acknowledgements:} 
I am especially indebted to P. Mattig and M. Mangano for informative
correspondence and discussions, as well as to B. Gary, S. Komamiya,
H. Neal, and M. Schmitt.   



\end{document}